\documentclass[aps,pre,twocolumn]{revtex4}  
  
\usepackage[utf8]{inputenc}
\usepackage{graphicx}
\usepackage{amssymb}
\usepackage{amsmath}
\usepackage{xcolor}

\begin{document}

\title{Critical patch size reduction by heterogeneous diffusion} 


\author{M. A. F. dos Santos$^{1}$, V. Dornelas$^{1,2}$, 
E. H. Colombo$^{3,4}$, and C. Anteneodo$^{1,5}$}

\address{$^{1}$ Department of Physics, PUC-Rio, Rua Marqu\^es de S\~ao Vicente 225, 22451-900, Rio de Janeiro, RJ, Brazil}
\address{$^{2} $ ICTP-SAIFR \& IFT-UNESP, Rua Dr. Bento Teobaldo Ferraz 271, 01140-070, São Paulo, SP, Brazil}
\address{$^{3}$ Department of Ecology \& Evolutionary Biology, Princeton University, Princeton, NJ 08544, USA\looseness=-1}
\address{$^{4}$ Department of Ecology, Evolution, and Natural Resources, Rutgers University, New Brunswick, NJ 08901, USA\looseness=-1}
\address{$^{5}$ Institute of Science and Technology for Complex Systems, Brazil}

\begin{abstract}
Population survival depends on a large set of factors that includes environment structure. Due to landscape heterogeneity, species can occupy   particular regions that provide  the ideal scenario for development, working as a refuge  from  harmful  environmental  conditions. Survival occurs if population growth overcomes the losses caused by 
adventurous individuals that cross the patch edge. In this work, we consider a single species dynamics in a bounded domain with a space-dependent diffusion coefficient. We investigate the impact of   heterogeneous diffusion  on the minimal patch size that allows population survival and show that, typically,  this critical size is smaller than  the one for  a homogeneous medium with the same  average diffusivity. 
\end{abstract}

\maketitle


\section{Introduction}

Species typically experience a patchy landscape, where only within certain regions individuals can find resources, shelter, and other key ingredients for survival~\cite{landscapeBook}. The landscape spatial structure shapes diverse macroscopic ecological patterns, affecting, for instance, the stability and diversity of ecosystems~\cite{HanskiBook,fahrig2003}.  
Particularly, the fragmentation and degradation of the habitats, accelerated by human activities, 
have been producing significant impacts on ecosystems, leading many species to extinction~\cite{maxwell2016biodiversity,congdon1993society}. 
Thus, it is, more than ever, a matter of interest to understand the role that habitat spatial features exert on species survival.

Focusing on a single patch, a central problem is to determine the critical patch size for species survival.  Typically, there exists a minimum size, $L_c$, that separates the extinction and survival regimes. Then, if the patch size $L$ is bigger than $L_c$, the population can grow, achieving a stationary profile at long time, while it goes extinct otherwise. The specific value of $L_c$ depends on the details of the environment and population dynamics.  

Pioneer investigations have addressed species survival assuming a time-independent bounded habitat and that individuals diffuse and reproduce with constant rates~\cite{kierstead1953size,skellam1951random,ludwing}. More recently, theoretical developments have been made to include demographic fluctuations, which arise from the stochastic character of the birth-death process~\cite{oasis}, and  experimental realization  using special strains of bacteria was performed to check the validity of the theory~\cite{perry}. Beyond this classical case, previous works have also discussed the effect of the spatio-temporal structure of the environment~\cite{cantrell2001a,convectionPRE,walkmask,periodicKenkre,colombo2016}, advection~\cite{pachepsky2005persistence,advection}, chemotaxis~\cite{Kenkre2008} and nonlinear  response~\cite{colombo2018}. These features affect the value of $L_c$, as they substantially modify the population dynamics at the edge of the habitat
~\cite{fagan1999,cantrell2002}. Furthermore, it has been shown that the belief that larger patches favor species survival fails if a strong nonlinearity is present~\cite{colombo2018}. Similarly, in the multi-species context, it has been shown that small patches can have high conservation value~\cite{Wintle2019}.

Despite previous works have already tackled the critical patch-size problem from many different perspectives, the effect of the space-dependent diffusion coefficient has not been sufficiently addressed.
Several mechanisms can make the diffusion coefficient depend on the particular location inside the patch.  For instance, the composition and structure of the medium through which individuals move can change approaching the patch edge. This is characteristic of the transition zone (ecotone) between habitat and non-habitat regions, which can distort animal movement~\cite{van1973home,ross2005edge}. Also, behavioral responses can affect mobility, as when individuals perceive at a distance~\cite{fagan2017} the drastic change in the environmental conditions near the edge of the habitat~\cite{fagan1999,cantrell2001,morales2002,colombo2015,bengfort2016}. Regardless of  the mechanisms that regulate the spatially-dependent diffusion coefficient, heterogeneities would affect the residence time of the organisms in the patch~\cite{vaccario2015}, thus impacting the critical patch size.

The role of space-dependent diffusion on the critical patch size has been studied before in simplified settings, assuming an abrupt change close to the edge of the patch~\cite{cantrell1999,gabriel2013}. 
 This  approximation  assumes a short-ranged response to the presence of the patch boundary and neglects the details about the spatial dependency of the movements. In this work, we extend this investigation for the case where the diffusion coefficient within the patch has a general form.

We consider a single species dynamics in a one-dimensional domain of size $L$, assuming that the diffusion coefficient depend on the position, $x$. At the microscopic-level, the random walk performed by each individual $i$ can be written in one-dimension as
\begin{eqnarray}
\dot{x}_i  = \sqrt{2D(x_i)}\;\eta_i(t),
\label{eq:ibm}
\end{eqnarray}
where $D(x)$ is the spatially-varying diffusion coefficient and  $\eta_i$  has typically a non-negligible correlation and probability density function that can range from Gaussian to L\'evy forms depending on the species and environmental conditions~\cite{levybacteria,bacporous,viswanathan2011physics,okubo2013diffusion}. In order to be able to employ an analytical treatment for the problem without disregarding the origins of the noise in Eq.~(\ref{eq:ibm}), we assume that $\eta$ is a Gaussian noise with very small (but non-null) correlation time~\cite{wong1965convergence,hanggi1995colored}. In this limit, independent on how the correlation decays in time, the corresponding macroscopic equation for the temporal evolution of the population density, $\rho(x,t)$, including diffusion and birth-death processes, 
can be written in one-dimension as
\begin{eqnarray}
\frac{\partial \ }{\partial t} \rho(x,t) =  \frac{\partial \ }{\partial x}  \sqrt{D(x)}  \frac{\partial \ }{\partial x}   \sqrt{D(x)}   \rho(x,t)+ f(\rho(x,t)), 
\label{eq:modeleq}
\end{eqnarray} 
 with   $x \in [-L/2,L/2]$ and Dirichlet boundary conditions. 
The first term in Eq.~(\ref{eq:modeleq}) introduces the space-dependent diffusion, which appears as prescribed by the Stratonovich interpretation \cite{stratonovich1966new,dos2018fractional,PhysRevE.94.032109,srokowski2009multiplicative,sandev2018heterogeneous}. The second term is a general growth rate which, in the present context, is only required to admit a Taylor expansion around the null population state.
The zero-density boundary condition $\rho(x=\pm L/2,t)=0$ mimics the harmful effects of the surroundings, which impose strong death rates, immediately killing  the  individuals that touch it. Although apparently drastic, this simplification has been useful in the context of homogeneous diffusion, and 
allows a first approach to the problem.

In Sec.~\ref{sec:results}
we derive an analytical expression to predict the critical patch size for  the problem described by Eq.~(\ref{eq:modeleq})  and provide illustrative examples for specific forms of $D(x)$. By comparing these results with the scenario in which the diffusion coefficient takes the average value inside the patch, $\Tilde{D} = \frac{1}{L} \int_{-L/2}^{L/2} D(x) dx$, we  show 
that heterogeneous diffusion  has a nontrivial effect in population survival.  We demonstrate that, under the Stratonovich interpretation, heterogeneous diffusion promotes the reduction of the critical patch size when contrasted to the averaged case where $D(x)=\tilde D$. For particular cases, including the rectangular, sinusoidal and stochastic diffusion profiles, we provide the explicit expression for critical patch size.
A mechanistic perspective on how heteregoeneous diffusion can emerge due to behavioral responses is also presented. 
At last, we show additional numerical results regarding the influence of the different stochastic interpretations of Eq.~(\ref{eq:ibm}) in Sec.~\ref{sec:formalism}.  
Sec.~\ref{sec:final} contains final remarks about the result.

\section{Critical patch size under space-dependent diffusion}
\label{sec:results}

The standard approach to obtain the critical patch size is based on the linear stability of the dynamics close to the extinction state, $\rho(x,t) =0$. In this regime, we Taylor expand the growth term in Eq.~(\ref{eq:modeleq}) up to first order. 
Noting that $f(0)=0$,  the remaining term is given by  $f(\rho)\simeq r \rho$, 
where $r=f'(0)$. 
Then, Eq.~(\ref{eq:modeleq}) becomes
\begin{eqnarray}
\frac{\partial \ }{\partial t} \rho(x,t) = \frac{\partial \ }{\partial x}  \sqrt{D(x)} \frac{\partial \ }{\partial x}  \sqrt{D(x)}\rho(x,t) + r \rho(x,t), \label{eq:modeleqlin}
\end{eqnarray}
 where $x\in[-L/2,L/2]$, with absorbing boundaries. 

 To circumvent the spatial dependency on $D$, we define~\cite{cherstvy2013anomalous}, 
\begin{eqnarray}
y(x)=\int^x dx'\frac{1}{\sqrt{D(x')}}, \label{newvariable0}
\end{eqnarray} 
which allows us to rewrite   Eq.~(\ref{eq:modeleqlin}) as 
\begin{eqnarray}
\frac{\partial\ }{\partial t} \overline{\rho}(y,t) =  \frac{\partial^2 \ }{\partial y^2} \overline{\rho}(y,t) +  r \overline{\rho}(y,t),\label{eq:diffusiony}
\end{eqnarray}
where $\overline{\rho}(y,t)=\sqrt{D(x)}\rho(x,t)$, and the new absorbing boundary condition is $\overline{\rho}(y(\pm L/2),t)=0$. 
Thus, in the new variable $y$, the problem reduces to that of the homogeneous diffusion treated in classical works~\cite{kierstead1953size,skellam1951random,ludwing,partial}, 
where individuals perform a standard state-independent Brownian motion.

\begin{figure}[b!]
    \centering
\includegraphics[width=0.98\columnwidth]{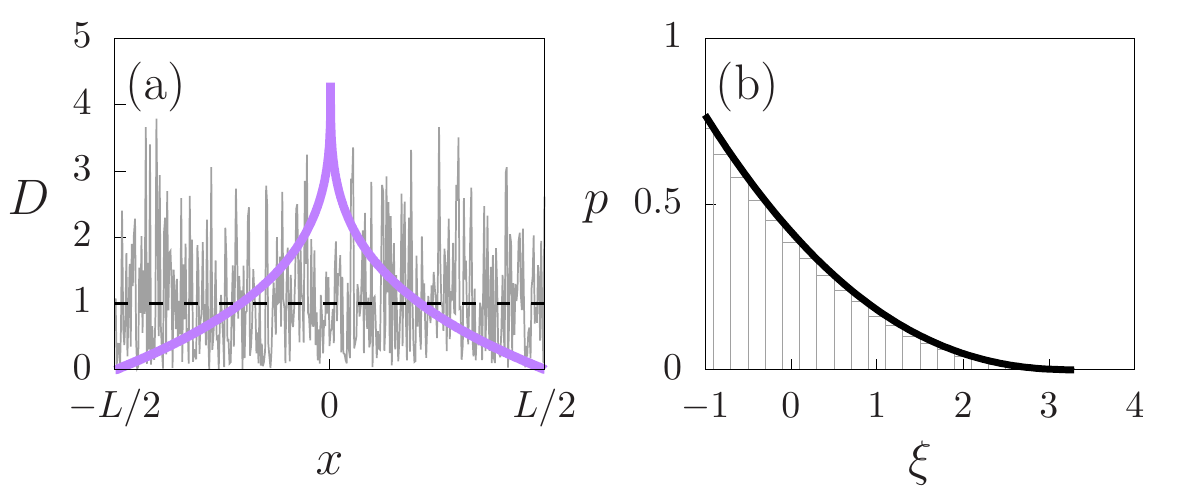}
    \caption{Diffusion coefficient profile and corresponding distribution.  (a) Diffusion coefficient (purple solid line) for a power-law profile, Eq.~(\ref{eq:power})  with $\alpha=0.3$, and the shuffled profile (gray solid line). (b) Distribution of the deviation from the average, $p(\xi)$, which is the same in both cases, thus generating identical result in Eq.~(\ref{eq:genLc}). Gray bars are from the numerical shuffling of $D(x)$ and the solid line the analytical result obtained from Eq.~(\ref{eq:PD}) (see Sec.~\ref{sec:powerlaw}).}
    \label{fig:shuffling}
\end{figure}

The population survives in the long-time if the extinction state is unstable, when the initial condition is non-null and contains a broad spectrum of modes. Solving Eq.~(\ref{eq:diffusiony}) by the method 
of separation of variables or through Fourier series, 
the contribution of each mode is   $\overline\rho_n(y,t)=e^{\lambda_n t}\cos(n \pi y/Y)$, where $Y=y(L/2)-y(-L/2)$ and $\lambda_n = r - (n \pi/Y)^2$, with $n=1,2,\ldots$. The population will grow in time if there is at least one mode $n$ with $\lambda_n>0$. Noting that $\lambda_1$ is the maximal rate, then it is clear that the condition for population survival is given $\lambda_1>0$. Otherwise, all other modes have negative growth rate. Therefore, the critical condition, $\lambda_1=0$, leads to
\begin{eqnarray} \label{eq:genLc}
Y_c= \int_{-L_c/2}^{L_c/2} \frac{1}{\sqrt{D(x)}}dx= \frac{\pi}{\sqrt{r}}.
\end{eqnarray}

For a homogeneous environment with  constant $D(x)=\tilde{D}$, 
the known expression $\tilde{L}_c\equiv\pi\sqrt{\frac{\tilde{D}}{r}}$ is recovered~\cite{partial}. 
The critical patch size arises from the balance between the flux that crosses the boundary and the growth inside the patch, as a consequence,  it increases with $\tilde{D}$ but decreases with $r$.

For the general heterogeneous case, let us consider the discretized version of  the integral in Eq.~(\ref{eq:genLc}), i.e., 
 $Y_c \simeq \frac{L_c}{N} \sum_{i=1}^N [D(x_i)]^{-1/2}$. 
 First notice  
that $Y_c$ remains the same by shuffling the values 
 of $D(x_i)$ within the integration interval. 
 In other words,  different profiles with the same
 distribution of values (see example in Fig.~\ref{fig:shuffling}) 
 yield the same result in Eq.~(\ref{eq:genLc}).
Mathematically, this is due to the fact that the integrand is a function of $D(x)$ only, which is a consequence of the homogeneity of the birth-death process contemplated in Eq.~(\ref{eq:modeleqlin}).

Furthermore, it is useful to write $D(x)=\tilde{D}[1+\xi(x)]$, 
such that $\langle \xi \rangle=0$, and $\xi>-1$ 
for the positivity of $D$, putting  
into evidence the variations $\xi$ around a 
reference level. 
We use this form into the discretized version of Eq.~(\ref{eq:genLc}), namely 
\begin{equation} \label{eq:LCdiscret}
  \frac{L_c}{N}  \sum_{i=1}^N (1+\xi_i)^{-1/2}    \simeq \tilde{L}_c \,,
\end{equation}
and search the extreme values of  
$h(\{\xi_i\})=\sum_{i=1}^N (1+\xi_i)^{-1/2}/N$, under the constraint $g(\{\xi_i\})=\sum_i^N \xi_i /N=0$. 
Through the method of Lagrange multipliers, 
we impose $\partial_{\xi_i} (g-\lambda h) =0$, 
obtaining $\xi_i=0$ for all $i$. From the analysis of the bordered Hessian, this corresponds to a minimum, 
with value  $h=1$. 
Then, from Eq.~(\ref{eq:LCdiscret}), in the continuum limit ($N\to\infty$), we get $L_c < \tilde{L}_c$.

Therefore, heterogeneous diffusion has the remarkable feature of typically producing a critical patch size smaller than in the corresponding averaged case, 
as we will see in  the examples  discussed in the following sections.

Noted that the distribution of values of the diffusion coefficient is the key feature, we focus on heterogeneities with distribution preserved  under changes of the size $L$.
In terms of  $D(x)$, this happens when the diffusivity depends on the position through the scaling $x/L$, that is $D(x)=\tilde{D}[1+\xi(2x/L)]$.  
In this case, Eq.~(\ref{eq:genLc}) becomes
\begin{eqnarray}
L_c=  
  \frac{ \tilde{L}_c }{ \frac{1}{2}\displaystyle \int_{-1}^{1}
 [1+\xi(z)]^{-\frac{1}{2}}dz}  \le \, \tilde L_c.
 \label{eq:criticalsizeG}
\end{eqnarray}
Performing a power-series expansion of the integrand around 
$\xi=0$, and taking into consideration that $\langle \xi \rangle=\int_{-1}^1\xi(z)dz=0$, 
at the lowest order, we obtain   
\begin{eqnarray}
  L_c    = 
\frac{\tilde L_c}{ 1 + \frac{3}{16} \displaystyle \int_{-1}^{1}[\xi(z)]^2 dz}  \le \, \tilde L_c,
\label{eq:ratioLcApprox}
\end{eqnarray}
with equality holding in the case of homogeneous diffusion, which clearly indicates that, for small variations, the critical patch size is smaller than that produced by the homogeneous environment with the same average diffusivity, and this deviation increases with the variability of $D$.


\subsection{Stochastic perspective} 

Since the main object that characterizes the heterogeneity and determines the value of the critical patch is the distribution $p(\xi)$, it is natural to develop a stochastic view of the diffusion profile, instead of thinking about the shape of $D(x)$. The following results provide a connection between these two perspectives, which will help to understand the particular cases tackled next.

We can interpret the deviation $\xi$ as a stochastic variable that assumes values in the interval $(-1,\infty)$ with a certain 
probability density function (PDF)  $p(\xi)$, which 
must verify $\langle \xi \rangle=\int_{-1}^\infty \xi \, p(\xi) \, d\xi =0$.
From this perspective, Eq.~(\ref{eq:criticalsizeG}) can be rewritten as 

\begin{eqnarray}\label{eq:LC}
L_c=  
  \frac{ \tilde{L}_c }{ \displaystyle \int_{-1}^\infty
 [1+\xi]^{-\frac{1}{2}} p(\xi) d \xi}.
 \label{eq:criticalsizeP}
\end{eqnarray}

This view allows to discuss the impact of disorder in terms 
of the PDF $p(\xi)$. 
Moreover, $p(\xi)$ can be related to a given profile of the diffusivity. 
Considering $z$ as uniform in 
the interval [-1,1], we make the change of variables $
\xi=\xi_i(z)$, for each interval $i$ where the function is 
monotonic, then  we have
\begin{equation} \label{eq:PD}
    p(\xi) =  \frac{1}{2} 
    \sum_i | d \xi_i/d z|^{-1} \,.
\end{equation}

\subsection{Rectangular profile}
\begin{figure}[t!]
\centering
\includegraphics[width=0.99\columnwidth]{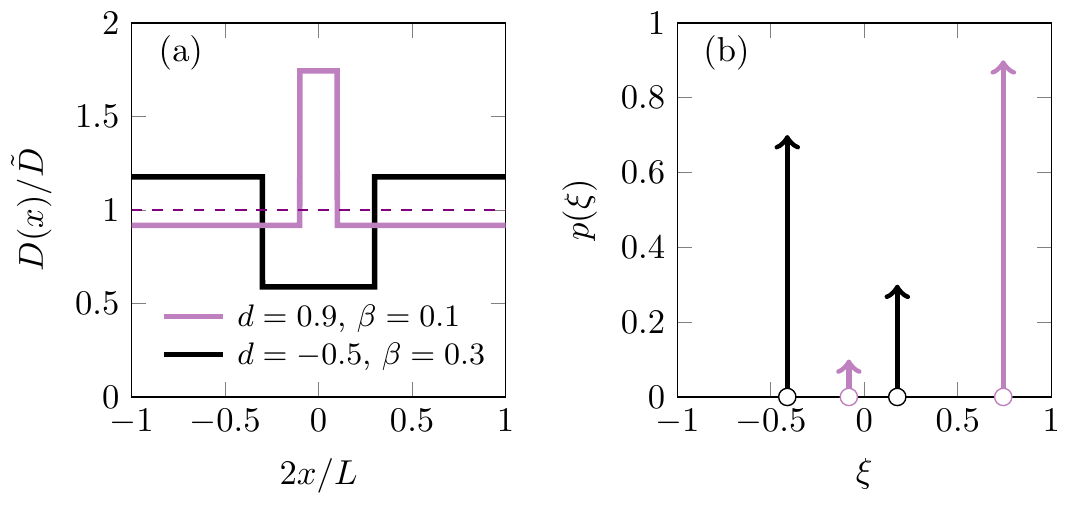}
\includegraphics[width=0.99\columnwidth]{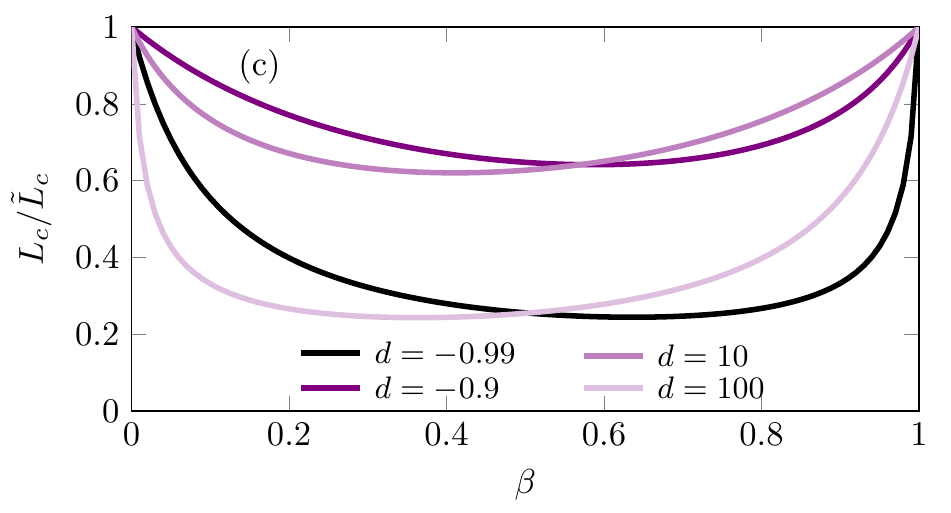}
\caption{ {\bf Rectangular.} 
Space-dependent diffusion coefficient scaled by its mean value $D(x)/\tilde{D} = 1 + \xi$ (a),   probability density function $p(\xi)$
 (b),   
critical patch size ratio $L_c/\tilde{L}_c$  vs. $\beta$ (c), 
given by Eq.~(\ref{eq:LCsquare}), such that $\beta L$ is the width of the central nucleus. In (b) the arrows represent Dirac delta functions. }
\label{fig:rectangular}
\end{figure}

Perhaps the simplest non-homogeneous case occurs when $D(x)$ assumes two values inside the patch of size $L$ instead of the single one in the homogeneous case. 
Let's say  a region of length $\beta L<L$ with 
coefficient $D_0 (1+d)$,  while the coefficient is $D_0$ otherwise, as illustrated in Fig.~\ref{fig:rectangular}a.
It can be defined through the Heaviside step function, $H$, as
\begin{equation}
D(x)= D_0 \left[1+d\, H \left(\beta  -\frac{2|x|}{L} \right) \right]\,,
\end{equation}
where $D_0$ is related to the mean value through
$\tilde{D}=D_0(1+d\beta)$,   $0<\beta<1$ and $d\beta>-1$ for positivity.

The associated PDF is   
$p(\xi) = \beta \delta(\xi+\frac{d\beta}{1+d\beta})
+[1-\beta]\delta(\xi-\frac{d(1-\beta)}{1+d\beta})$, 
the sum of two Dirac delta functions (see 
Fig.~\ref{fig:rectangular}b).

From Eq.~(\ref{eq:criticalsizeG}), we obtain the 
explicit expression for the critical size (wee Fig.~\ref{fig:rectangular}b)
\begin{equation} \label{eq:LCsquare}
L_c/\tilde{L}_c=\frac{\sqrt{1+d}}{\sqrt{1+  \beta d}
\left(  (1-\beta)\sqrt{1+d} + \beta \right)    } \leq 1\,.
\end{equation}
It does not depend on the localization of the nucleus, which can be shifted from the origin, or even fragmented 
in many nuclei, of total size $\beta L$. It only depends on the 
proportion of the patch, $\beta$, adopting either of the two values.

When $d \to -1$, there is a region with null diffusivity, 
therefore the population grows without diffusive losses and as a consequence, the critical size tends to zero. 
The same occurs in the opposite limit when there is a region with very high relative diffusivity ($d\to\infty$) in comparison to that of the remaining habitat. 
The homogeneous case occurs when $d=0$, or $\beta=0$ or 1, and it requires the maximal patch size for survival. 
Moreover, we can see in Fig.~\ref{fig:rectangular}b that there is an optimal value of $\beta$ that minimizes $L_c/\tilde L_c$.  Also note the high 
contrast between heterogeneous and homogeneous diffusion, when $d>>1$ or $d\simeq -1$, yielding a reduction of $75\%$ of the critical size in the cases shown.

\subsection{Sinusoidal profile} 
 
\begin{figure}[t!]
\centering
\includegraphics[width=0.99\columnwidth]{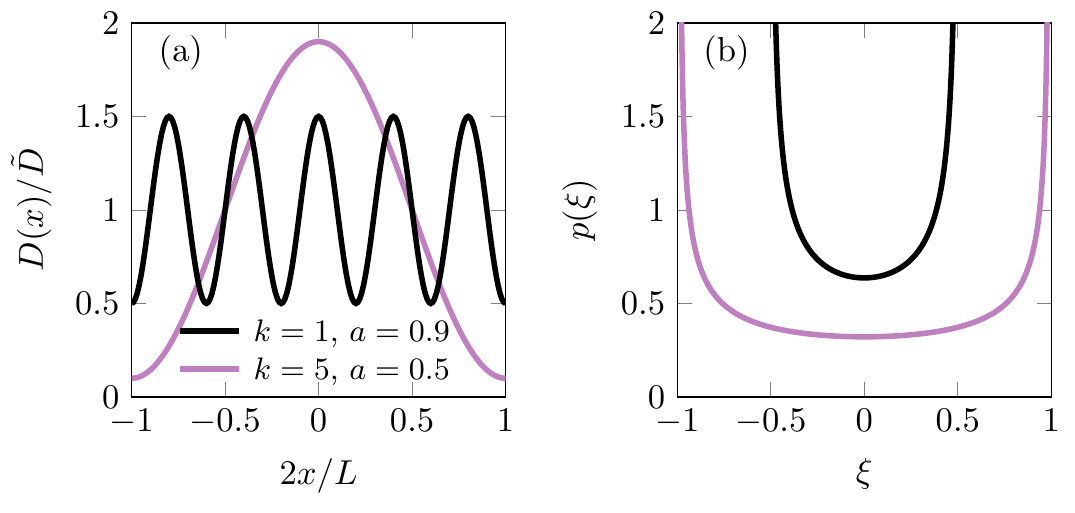}
\includegraphics[width=0.99\columnwidth]{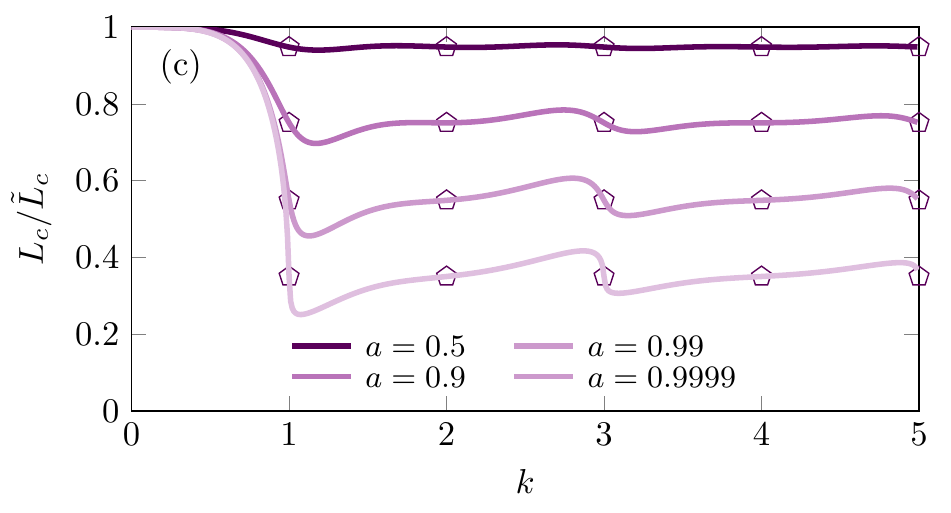}
\caption{ {\bf Sinusoidal.} 
Same as in Fig.~\ref{fig:rectangular} for the 
sinusoidal diffusivity, with $\phi=0$. 
In (b), the PDF corresponds to any integer $k$. 
In (c) the symbols highlight that the critical ratio 
remains invariant for integer values of $k$, even 
if it were $\phi \neq 0$.
}
\label{fig:sin}
\end{figure}

Another important case is when the variation around the 
mean value of the diffusion coefficient is sinusoidal (Fig.~\ref{fig:sin}a), 
that is
\begin{equation}
D(x)= D_0\left[ 1 + a\cos \left(  \frac{2k \pi x}{L}  + \phi \right)\right]\,.
\end{equation}

When an integer number  of periods 
fits the patch (i.e., $k \in Z$),  
$\tilde{D}=D_0$,  and  the result for the ratio $L_c/\tilde{L}_c$ does not 
depend on the phase constant $\phi$, nor in the 
periodicity given by $k$, because the distribution of values remains unchanged.
In fact, Eq.~(\ref{eq:PD}) yields
 $p(\xi)=1/(\pi \sqrt{a^2- \xi^2})$ 
 for $ |\xi|<a<1$, which only depends on the amplitude $a$ (see Fig.~\ref{fig:rectangular}c).
Differently, when $k$ is noninteger, 
$\tilde{D}=D_0[1 + a\cos\phi \sin( k \pi)/(k\pi)]$  and the ratio of critical sizes  
depends on $k$ as well as on the phase $\phi$ 
(see Fig.~\ref{fig:sin}c). 
In particular, for integer $k$ and  small amplitude, Eq.~(\ref{eq:ratioLcApprox}) predicts
$L_c/\tilde{L}_c \simeq 1/[1+ (3/16) a^2]$.

\subsection{Power-law profile}
\label{sec:powerlaw}

Let us consider the  {\it power-law} function 

\begin{equation} \label{eq:power}
D(x)= D_0\left(1-d\left|\frac{2x}{L} \right|^{\alpha}\right)\,,
\end{equation} 
where $D_0=\tilde{D}(\alpha +1)/(\alpha+1-d)$. 
It leads to the ratio of critical sizes 
\begin{equation} \label{eq:Lcpower}
L_c/\tilde{L}_c =   
\left[
\sqrt{1-\frac{d}{\alpha +1} } \, \,_2F_1\left(\frac{1}{2},\frac{1}{\alpha },1+\frac{1}{\alpha },d \right) \right]^{-1} 
\le 1, 
\end{equation} 
which is unity in the limit $\alpha \to \infty$. 
 
\begin{figure}[h!]
\centering
\includegraphics[width=0.99\columnwidth]{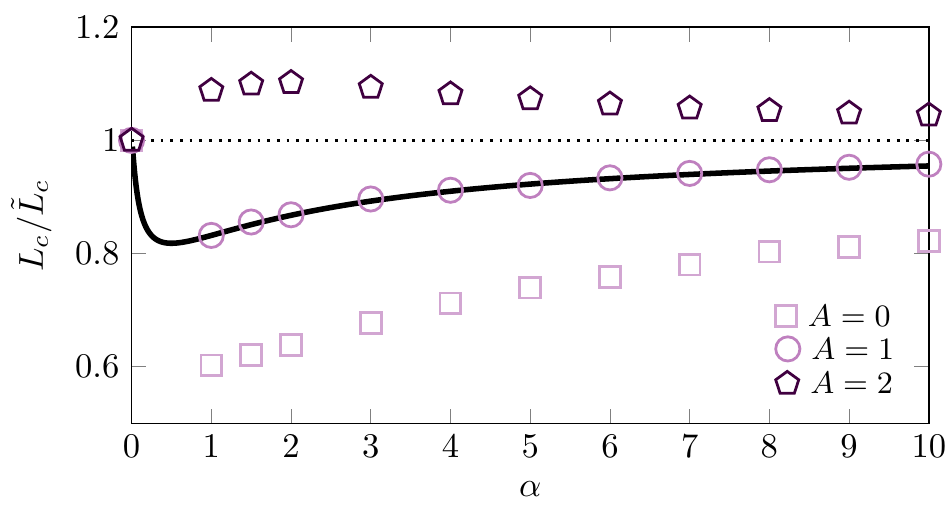}
\caption{Critical size ratio for 
the  power-law case defined in Eq.~(\ref{eq:power}), 
with $d=0.96$ (solid line), following Eq.~(\ref{eq:Lcpower}).
Symbols correspond to numerical simulations for 
different values of the prescription parameter $A$.
In the limits $\alpha\to 0$ and $\alpha\to \infty$, 
we have $\tilde{L}_c/L_c \to 1$ in all 
cases, because $D(x)\to \tilde{D}$. 
The horizontal dotted line highlights the unity ratio. }
\label{fig:A}
\end{figure}

For the particular case  $d=1$, $D(x)$ vanishes at the boundaries. 
In this case, 
$\xi(z)=[1-  (\alpha+1)|z|^\alpha]/\alpha$, 
with probability 
$p(\xi)= (1-\alpha\xi)^{1/\alpha-1}/(\alpha+1)^{1/\alpha}$ in $[-1,1/\alpha]$.
When $\alpha=1$ (triangular profile), it  corresponds to the uniform 
distribution in $[-1,1]$. 
The limit $\alpha \to 0$, yields the anomalous case 
$D(x)=-\tilde{D}\ln|2x/L|$, corresponding to the exponential $p(\xi)=\exp(-\xi-1)$. 
However, for $d \neq 1$, in the limit $\alpha \to 0$, we also 
recover the
homogeneous case, as in the limit $\alpha \to \infty$. 
In Fig.~\ref{fig:A}, we plot the ratio of critical sizes vs 
$\alpha$, for  a concave power-law profile, $d=0.96>0$ (solid line).

\subsection{Edge-response profile}
\label{sec:nonscaling}

When modeling the mechanisms responsible for triggering the spatial dependency on individuals' mobility, the scaling dependence might not be suitable. But Eq.~(\ref{eq:genLc}) can be used in a much broader scenario. Then, to spark possible ideas in this sense, let us discuss the case where mobility is a function of the distance from the patch edge. This can reflect changes in the landscape structure and composition during the transition from the habitat and non-habitat regions  known as \textit{ecotone}. Alternatively, it can mimic  the behavioral changes as individuals perceive the patch boundary, instead of perceiving that the medium is changing~\cite{fagan2017}.

Assuming that individuals' mobility is reduced near the boundary, both views lead to a diffusion coefficient of the form of
$
D(x)=D_0[1 - \gamma(|x-L/2|)]
$
where $\gamma$ is a function that  
is $\gamma(0)=1$  and vanishes far from the boundary, such that the diffusion coefficient attains its maximum value $D_0$.

\begin{figure}[t!]
    \centering
    \includegraphics[width=\columnwidth]{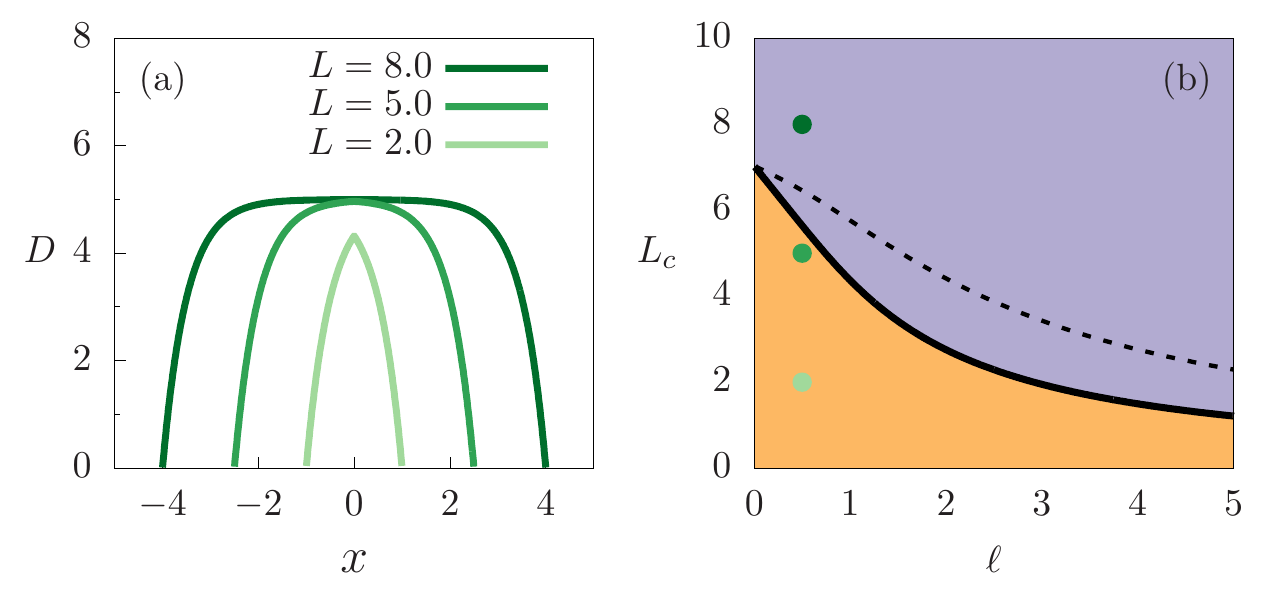}
    \caption{Diffusion coefficient (a) and critical patch size (b) for an edge distance-dependent mechanism with the form $D(x)=D_0(1-\exp[(|x|-L/2)/\ell])$ with  $D_0=5.0$ and $\ell=0.5$. In panel (b), for $r=1$, the orange and lilacs regions correspond to the survival and extinction phases, respectively. The dashed line represents $\tilde{L}_c$. Greenish dots correspond to the cases in panel (a).}
    \label{fig:mec}
\end{figure}

A simple case which suits this scenario is given by $D(x)=D_0(1-\exp[(|x|-L/2)/\ell])$, where $\ell$ is the characteristic scale of the response to the edge. 
An illustration of this diffusive profile is depicted in Fig.~\ref{fig:mec}a  and the critical patch size as a function of the response scale $\ell$ in Fig.~\ref{fig:mec}b. 
 Note that the profile shape is not preserved as the patch size increases (Fig.~\ref{fig:mec}a). Hence,   Eq.~(\ref{eq:criticalsizeG}) cannot be applied, but Eq.~(\ref{eq:genLc}) yields the closed form, 
$L_c=4\ell\ln[\cosh(\pi\sqrt{D_0/r}/(4\ell))]$  (Fig.~\ref{fig:mec}b). 
In the limit $\ell\to 0$,  the profile converges to the homogeneous case, which produces $L_c=\tilde L_c$. As $\ell$ increases, $L_c$ decreases, vanishing at $\ell\to \infty$. In the examples of Fig.~\ref{fig:mec}a (with $\ell=0.5$), $L_c\simeq 5.64$, as a consequence, for $L=2.0$ and $5.0$, the population goes extinct, while for $L=8.0$ it survives, as indicated by the dots with correspondent colors in Fig.~\ref{fig:mec}b. 
Although the critical patch size for the average value of $D$, $\tilde L_c$, has no closed form, we numerically checked that  the statement $L/\tilde L_c\leq 1$ remains valid (dotted line in Fig.~\ref{fig:mec}b).

\section{Other interpretations of the space-dependent diffusion}
\label{sec:formalism}

 The derivation of  the macroscopic term for the diffusion processes from the stochastic individual-level, 
 is not  unique.  
There are different versions of a space-dependent diffusion equation \cite{hanggi1982nonlinear,klimontovich1990ito,ito1944,stratonovich1966new,wang2020anomalous} and all these forms converge to the standard case when the diffusion coefficient is a constant in space and time. 
For all versions, the probability density $\rho(x,t)$ depends on the particular form of $D(x)$. A general class of heterogeneous-diffusion equations is
 \begin{eqnarray}
 \frac{\partial \ }{\partial t} \rho(x,t) =  \frac{\partial \ }{\partial x} \left\{  D(x)^{1-\frac{A}{2}}  \frac{\partial \ }{\partial x}  \left[ D(x)^{\frac{A}{2}} \rho(x,t) \right] \right\},
 \label{eq:generalDIFFUSION}
 \end{eqnarray}
 in which $A$ is, in principle, a positive real number  ($A=1$ in our case). For $A\in  \{0,1,2 \}$, it  defines the heterogeneous diffusion equation accordingly to different well-known prescriptions:   Hänggi-Klimontovich  ($A=0$),  Stratonovich ($A=1$),  and It\^o ($A=2$) formalisms.
 A possible underlying stochastic dynamics 
 associated to Eq.~(\ref{eq:generalDIFFUSION})
 is given by Eq.~(\ref{eq:ibm}) 
 accompanied by the interpretation for the noise $\eta$ associated to the value of $A$. 
  Alternatively, we can adopt, for instance, the Stratonovich prescription, and modify 
  the stochastic equation adding 
  a drift term associated to the chosen value of $A$, yielding $\dot{x}_i  =(1-A)D'(x_i)/2+\sqrt{2D(x_i)}\eta_i(t)$.
 
 This general form can be used to access the consequences of each interpretation in relevant macroscopic outcomes. 
For instance, recently, the general class of diffusive process  in Eq.~(\ref{eq:generalDIFFUSION})  has been used to investigate the impact of each prescription in the normalization of the probability distribution of a particle diffusing in a heterogeneous environment with $D(x)\propto |x|^\beta$~\cite{PhysRevE.99.042138}. In this section, we address the role of the different interpretations of Eq.~(\ref{eq:ibm}) on the critical patch size.

Rather than entering the interpretation  dilemma, it may be more valuable to understand the origins and dynamics responsible for the noise in question, which will naturally lead to the appropriate interpretation~\cite{kupferman2004ito}. 
In our case, we have adopted the Stratonovich interpretation ($A=1$),  as mentioned in the introduction,  implicitly assuming that the noise, $\eta$, in Eq.~(\ref{eq:ibm}) has a temporal correlation much longer than the relaxation time promoted by the inertia  of the particles (individuals). Note that this consideration precedes Eq.~(\ref{eq:ibm}), for which the over-damped limit has already been taken. 
More generally, depending on the microscopic details of the walk performed by the individuals, different values of $A$, even fractional ones
~\cite{tupikina2020}, might be appropriate. 
 For instance,  when the  particle dynamics  relaxation time and the noise temporal correlation vanish,   with the former  surpassing the latter one, the It\^o interpretation ($A=2$) is the one that naturally emerges~\cite{kupferman2004ito}.

In Fig.~\ref{fig:A}, we compare the outcomes for different values of $A$. 
To do that, we numerically integrated Eq.~(\ref{eq:modeleq}) using the generalized diffusion term in Eq.~(\ref{eq:generalDIFFUSION}) starting from the null homogeneous state plus a positive small random noise. We applied a standard forward-time-centered-space scheme which is fourth-order Runge-Kutta in time and second-order in space, with discrete-time step $\Delta t = 10^{-5}$ and cell size $\Delta x = 0.01$.

We observe that the ratio $L_c/\tilde L_c$ has significantly different values that  increase with $A$. 
These results can be understood in light of results on the mean first passage time under heterogeneous diffusion. To do that,  first  recall that the critical patch size is achieved from the balance between diffusive losses at the borders and growth in the habitat. 
In other words,  this occurs when the individuals' habitat residence time, $\tau_h$, equals the reproduction inter-event time, $\tau_r$,  i.e., individuals reproduce exactly once before hitting the boundary and dying. For the homogeneous diffusion case, in which all interpretations produce identical values, $\tau_h \sim L^2/D$ and $\tau_r \sim 1/r$, which leads to $L_c \sim \sqrt{D/r}$
~\cite{perry}. Under heterogeneous diffusion, it has been shown that $\tau_h$  is significantly affected by $A$~\cite{vaccario2015}. In the case discussed here, $\tau_h$ increases with $A$,  leading to a larger critical patch size as $A$ increases. This picture, however, can change depending on the particular problem treated~\cite{vaccario2015}. 

 Lastly, note  the result obtained from Eq.~(\ref{eq:LCdiscret}), that $L_c/\tilde L_c \le 1$ for any $D(x)$, was derived for the diffusion equation associated to $A=1$ and it is not expected to apply for any $A$. 
In fact, for the It\^o case, $A=2$, in Fig.~\ref{fig:A},   $L_c/\tilde L_c > 1$.

\section{Final remarks}
\label{sec:final}

We have shown that space-dependent diffusion, which appears as prescribed by the Stratonovich interpretation, typically favors survival, by reducing the critical patch size 
in population dynamics described by Eq.~(\ref{eq:modeleqlin}). 

We noted that Eq.~(\ref{eq:genLc}) is not affected by shuffling the values of $D(x)$, 
which allowed the analysis from the perspective of the distribution of values around the mean, 
$p(\xi)$.  
However, it is important to comment that the presence of any correlation between population growth and diffusion, such as a density-dependent diffusion coefficient~\cite{colombo2018}, would change the form of Eq.~(\ref{eq:genLc})  in such a way that the specific location and values of $D$ would matter.

Assuming that the type of heterogeneity present, characterized by $p(\xi)$, is kept invariant, we investigated the cases in which the profile of the diffusion coefficient scales with the habitat size. This allowed us to extract simple expressions to show how heterogeneous diffusion affects the critical habitat size in comparison to the average level.  
We also provided an illustration of the nonscaling case, which similarly reduces the critical size, in accord with Eq.~(\ref{eq:LCdiscret}).  

Furthermore, we considered a generalized form of heterogeneous diffusion that includes different interpretations of the underlying stochastic dynamics. 
We observed that in contrast to the Stratonovich case  ($A=1$),   heterogeneous diffusion can lead to  an  increase of the habitat size under the It\^o interpretation 
($A=2$).

 All these results highlight that the details of how individual behavior and  spatial structure of the environment change inside patch boundaries should be taken into account in ecological management and natural reserve (refuge) design~\cite{cantrell1999}. This adds to the point that neglecting the   internal variability can lead to incorrect predictions about the macroscopic behavior of the system, a fact that has already been remarked in other ecological contexts~\cite{allen2001,colombo2019}. 

Let us also comment, that although we made our study for a one-dimensional setting, similar qualitative results are expected in two dimensions. The investigation of the individual  particle  residence time in higher dimensions might provide insights in this sense~\cite{vaccario2015,metzler}.

As a perspective for future work, it might be interesting to study the corresponding equation with density-dependent growth, e.g., with   a power-law dependence which is nonlinearizable~\cite{colombo2018}, 
or with a  birth-death process which is  also  spatially dependent~\cite{cantrell2001}. This would break-down the shuffling statement (Fig.~\ref{fig:shuffling}) bringing the shape of the profile to the spotlight. This would probably require a numerical approach since only a few cases may be analytically accessible.

\section*{Acknowledgements} 
We thank Ra\'ul O. Vallejos for 
helpful suggestions. 
MAFS, VD and CA acknowledge partial financial support by the 
Coordena\c c\~ao de Aperfei\c coamento de Pessoal de N\'{\i}vel Superior
 - Brazil (CAPES) - Finance Code 001. CA also acknowledges partial support by 
 Conselho Nacional de Desenvolvimento Cient\'{\i}fico e Tecnol\'ogico (CNPq) and Fundação de Amparo à Pesquisa do Rio de Janeiro (FAPERJ).
VD also acknowledges partial support by FAPESP through the ICTP-SAIFR grant 2016/01343-7 and Postdoctoral fellowship 2020/04751-4.


 \end{document}